\documentclass[conference]{IEEEtran}

\IEEEoverridecommandlockouts                              

\usepackage[overload]{empheq} 
\usepackage{cite}
\usepackage{graphicx}
\usepackage{epstopdf}
\usepackage{graphics} 
\usepackage{epsfig} 
\usepackage{color}
\usepackage{amsmath}
\usepackage{amsthm}
\usepackage{amssymb} 
\usepackage{verbatim}  
\usepackage{mathtools}
\usepackage{subfig}
\usepackage{tabularx}
\usepackage{amsthm}
\usepackage{cases}
\usepackage{verbatim}  
\usepackage{relsize}
\newtheorem{theorem}{Theorem}

\newtheorem{definition}{Definition}
\newtheorem{remark}{Remark}
\usepackage[ colorlinks = true,
linkcolor = blue,
urlcolor = blue,
citecolor = red,
anchorcolor = green,
]{hyperref}

\def\mcl{\mathcal}
\newcommand{\be}{\begin{equation}}
\newcommand{\ee}{\end{equation}}

\allowdisplaybreaks[2]

\title{A Tunable Measure for Information Leakage}
\author{\IEEEauthorblockN{Jiachun Liao, Oliver Kosut, Lalitha Sankar}
	\IEEEauthorblockA{School of Electrical, Computer and Energy Engineering,\\
		Arizona State University\\
		Email: \{jiachun.liao,lalithasankar,okosut\}@asu.edu}
	\and
	\IEEEauthorblockN{Flavio P. Calmon}
	\IEEEauthorblockA{School of Engineering and Applied Sciences\\
		Harvard University\\
		Email: fcalmon@g.harvard.edu\\
	}
\thanks{This material is based upon work supported by the National Science Foundation under Grant No. CCF\--1350914.}
}

\begin{document}
	\maketitle	
	
\begin{abstract}
	A tunable measure for information leakage called \textit{maximal $\alpha$-leakage} is introduced. This measure quantifies the maximal gain of an adversary in refining a tilted version of its prior belief of any (potentially random) function of a dataset conditioning on a disclosed dataset. The choice of $\alpha$ determines the specific adversarial action ranging from refining a belief for $\alpha =1$ to guessing the best posterior for $\alpha = \infty$, and for these extremal values this measure simplifies to mutual information (MI) and maximal leakage (MaxL), respectively.  For all other $\alpha$ this measure is shown to be the Arimoto channel capacity. Several properties of this measure are proven including: (i) quasi-convexity in the mapping between the original and disclosed datasets; (ii) data processing inequalities; and (iii) a composition property. 
	

\end{abstract}

\section{Introduction}

Information leakage metrics seek to quantify an adversary's ability of inferring information about one quantity from another.
Mutual information (MI) is a classic measure for quantifying information and often used to measure information secrecy \cite{SecrecySystem_Shannon49} or leakage in data publishing settings
\cite{sankar_utility-privacy_2013,calmon2014allerton}.  More recently, Issa \textit{et al.} introduced a measure, called \textit{maximal leakage} (MaxL), for a guessing adversary that quantifies the maximal multiplicative gain of an adversary, with access to a disclosed dataset, to guess \emph{any} (\emph{possible random}) \emph{function} of the original dataset \cite{IssaKW16}.

Information leakage measures can be viewed through the lens of adversarial inference capabilities, and therefore, quantified via a loss function that the adversary seeks to minimize. The choice of a loss function provides a concrete measure of the gain in adversarial inference capability. For example, the definition of MaxL
can be interpreted in terms of an adversary seeking to minimize the 0-1 loss function, which induces the adversary towards a hard decision, i.e., a maximum likelihood estimator.
On the other hand, when MI is used as a leakage measure, the underlying loss function is the \emph{logarithmic loss} (log-loss) function \cite{Merhav1998,Courtade2011,Calmon_privacy_2012}, which models a (soft decision) belief-refining adversary. These two models capture two extremal actions of adversaries.
Can these measures be viewed through the same framework? In this paper, we introduce a tunable measure, called \textit{maximal $\alpha$-leakage}, for information leakages, which encompasses MI (for $\alpha=1$) and MaxL (for $\alpha=\infty$) and allows continuous interpolation between the two extremes. The parameter $\alpha$ can be viewed as a tunable parameter that determines how much weight the adversary gives to its posterior belief.

In this paper, we define two tunable measures for information leakages in Section \ref{Sec:Information Leakage Measures}: $\alpha$-leakage (Definition \ref{Def:alphaLeakge}) and maximal $\alpha$-leakage (Definition \ref{Def:GeneralLeakge}). In Section \ref{Sec:Information Leakage Measures}, we prove that the $\alpha$-leakage can be expressed as Arimoto mutual information (A-MI) (Theorem \ref{Thm:DefEquialentExpression_alphaleakage}), and the maximal $\alpha$-leakage is equivalent to the supremum of A-MI and Sibson mutual information (S-MI) (Theorem \ref{Thm:DefEquialentExpression}) over all distributions of the original dataset. In Section \ref{Sec:Properties}, we prove several important properties of the maximal $\alpha$-leakage.

\section{Preliminaries}\label{Sec:Preliminaries}
We begin by reviewing R{\'e}nyi entropy and divergence \cite{measures_renyi1961}.
\begin{definition}
		Given a discrete distribution $P_X$ over a finite alphabet $\mcl X$, the R{\'e}nyi entropy of order $\alpha\in (0,1)\cup(1,\infty)$ is defined as
		\begin{align}
		\label{eq:renyi_entropy}
		H_{\alpha}(P_X)=
		\frac{\alpha}{1-\alpha}\log\|P_X\|_{\alpha}. 
		\end{align}
		Let $Q_X$ be a discrete distribution over $\mcl X$. The R{\'e}nyi divergence (between $P_X$ and $Q_X$) of order $\alpha\in (0,1)\cup(1,\infty)$ is defined as 
		\begin{align}
		\label{eq:renyi_divergence}
		D_{\alpha}(P_X\|Q_X)=\frac{1}{\alpha-1}
		\log\left(\sum\limits_{x}\frac{P_X(x)^{\alpha}}{Q_X(x)^{\alpha-1}}\right).
		\end{align}
		Both of the two quantities are defined by their continuous extension for $\alpha=1$ or $\infty$.
\end{definition}   
 The $\alpha$-leakage and max $\alpha$-leakage metrics can be expressed in terms of Sibson mutual information (S-MI) \cite{alphaMI_Sibson1969} and Arimoto mutual information (A-MI) \cite{AlphaMI_Arimoto1975}. These quantities generalize the usual notion of MI. We review these definitions next.
 \begin{definition}
 	Let discrete random variables $(X,Y)\sim P_{XY}$ with $P_X$ and $P_Y$ as the marginal distributions, respectively, and $Q_Y$ be an arbitrary marginal distribution of $Y$. The Sibson mutual information (S-MI) of order $\alpha\in(0,1)\cup(1,\infty)$ is defined as
 	\begin{align}
 	\label{eq:Def_SibsionMI}
 	I_\alpha^{\text{S}}(X;Y)&\triangleq \inf_{Q_Y}\,D_\alpha(P_{XY}\|P_X\times Q_Y),\\
 	\label{eq:Sibson_MI}
 	&= \frac{\alpha}{\alpha-1}\log \sum\limits_{y}\left(\sum\limits_{x}P_X(x)P_{Y|X}(y|x)^{\alpha}\right)^{\frac{1}{\alpha}}
 	\end{align}
 	The Arimoto mutual information (A-MI) of order $\alpha\in(0,1)\cup(1,\infty)$ is defined as
 	\begin{align}
 	\label{eq:Def_ArimotoMI}
 	I_\alpha^{\text{A}}(X;Y)&\triangleq H_{\alpha}(X)-H_{\alpha}(X|Y)\\
 	\label{eq:Arimoto_MI}
 	&=\frac{\alpha}{\alpha-1}\log\sum\limits_{y }\mathsmaller{\left(\frac{\sum\limits_{x}P_X(x)^{\alpha}P_{Y|X}(y|x)^{\alpha}}{\sum\limits_{x }P_X(x)^{\alpha}}\right)^{\frac{1}{\alpha}}},
 	\end{align}
 	where $H_{\alpha}(X|Y)$ is Arimoto conditional entropy of $X$ given $Y$ defined as
 	\begin{align}
 	\label{eq:Def_ArimotoConditionalEntropy}
 	H_{\alpha}(X|Y)=
 	\frac{\alpha}{1-\alpha}\log\sum\limits_{y}\mathsmaller{\left(\sum\limits_{x}\mathsmaller{P_X(x)^{\alpha}P_{Y|X}(y|x)^{\alpha}}\right)^{\frac{1}{\alpha}}}.
 	\end{align}
 	All of these quantities are defined by their continuous extension for $\alpha=1$ or $\infty$.
 \end{definition}

\section{Information Leakage Measures} \label{Sec:Information Leakage Measures}
In this section, we formally define the tunable leakage measures: $\alpha$-leakage and maximal $\alpha$-leakage.

Let $X$, $Y$ and $U$ be three discrete random variables with finite supports $\mcl X$, $\mcl Y$ and $\mcl U$, respectively. Let $\hat{X}$ be an estimator of $X$ and $P_{\hat{X}|Y}$ indicate a strategy for estimating $X$ given $Y$.
We denote the probability of correctly estimating $X=x$ given $Y=y$ as
\begin{align}
\label{eq:Notation_ProbCorrectEst}
P_c(P_{\hat{X}|Y},x,y)=P_{\hat{X}|Y}(x|y)=\mathbb{P}(\hat{X}=x|x,y).
\end{align}
Let $X$ and $Y$ represent the original data and disclosed data, respectively, and let $U$ represent an arbitrary (potentially random) function of $X$ that the adversary (a curious or malicious user of the disclosed data $Y$) is interested in learning. In \cite{MaximalLeakage_Issa2016}, Issa \textit{et al.} 
introduced MaxL to qualify the maximal gain in an adversary's ability of guessing $U$ by knowing $Y$. We review the definition below.
\begin{definition}[{\cite[Def. 1]{MaximalLeakage_Issa2016}}]\label{Def:MaximalLeakage}
	Given a joint distribution $P_{XY}$, the \textit{maximal leakage} from $X$ to $Y$ is
	\begin{equation}\label{ml_op_def}
		\mcl L_{\text{MaxL}}(X\to Y)\triangleq\sup_{U- X- Y} \log \frac{\max\limits_{u} \mathbb{E}\left[\mathbb{P}(\hat{U}=u|Y)\right]}{\max\limits_{u} \mathbb{P}(\tilde{U}=u)}.
	\end{equation}
		where both estimators $\hat{U}$ and $\tilde{U}$ take values from the same arbitrary finite support as $U$.
\end{definition} 
\begin{remark}\label{Remark:MaxL}
	Note that from \eqref{eq:Notation_ProbCorrectEst}, the numerator of the logarithmic term in \eqref{ml_op_def} can be explicitly written as
		\begin{align}
			\max\limits_{u}\mathbb{E}\left[\mathbb{P}(\hat{U}=u|Y)\right]=\max\limits_{u}\sum\limits_{y}P_{Y}(y)P_{\hat{U}|Y}(u|y).
		\end{align}
	In Definition \ref{Def:MaximalLeakage}, $U$ represents any (possibly random) function of $X$. The numerator represents the maximal probability of correctly guessing $U$ based on $Y$, while the denominator represents the maximal probability of correctly guessing $U$ \emph{without} knowing $Y$. Thus, MaxL quantifies the maximal gain (in bits) in guessing any possible function of $X$ when an adversary has access to $Y$.
\end{remark}

We now present $\alpha$-leakage and maximal $\alpha$-leakage (under the assumptions of discrete random variables and finite supports). 
The $\alpha$-leakage measures \textit{various} aspects of the leakage (ranging from the probability of correctly guessing to the posteriori distribution) about data $X$ from the disclosed $Y$.  
\begin{definition}[{$\alpha$-Leakage}]\label{Def:alphaLeakge}
	Given a joint distribution $P_{XY}$ and an estimator $\hat{X}$ with the same support as $X$, the $\alpha$-leakage from $X$ to $Y$ is defined as
	\begin{align}
	\label{eq:alphaLeak_definition}
	\mathsmaller{\mcl L_{\alpha}(X\hspace{-0.04in}\to \hspace{-0.04in}Y)
	\triangleq\frac{\alpha}{\alpha-1}\log\frac{\max\limits_{P_{\hat{X}|Y}}\mathbb{E}\left[\mathbb{P}(\hat{X}=X|X,Y)^{\frac{\alpha-1}{\alpha}}\right]}{\max\limits_{P_{\hat{X}}}\mathbb{E}\left[\mathbb{P}(\hat{X}=X|X)^{\frac{\alpha-1}{\alpha}}\right]}}
	\end{align}
for $\alpha\in(1,\infty)$ and by the continuous extension of \eqref{eq:alphaLeak_definition} for $\alpha = 1$ and $\infty$.
\end{definition}
From \eqref{eq:Notation_ProbCorrectEst}, the numerator of the logarithmic term in \eqref{eq:alphaLeak_definition} can be explicitly written as
\begin{align}
&\max\limits_{P_{\hat{X}|Y}}\mathbb{E}\left[\mathbb{P}(\hat{X}=X|X,Y)^{\frac{\alpha-1}{\alpha}}\right]\nonumber\\
=&\max\limits_{P_{\hat{X}|Y}}\sum\limits_{xy}P_{XY}(xy)P_{\hat{X}|Y}(x|y)^{\frac{\alpha-1}{\alpha}}.
\end{align}
Analogous to the analysis for MaxL in Remark \ref{Remark:MaxL}, $\alpha$-leakage quantifies the multiplicative increase in the expected reward for correctly inferring $X$ when an adversary has access to $Y$. 

Whereas $\alpha$-leakage captures how much an adversary can learn about $X$ from $Y$, we also wish to quantify the information leaked about \textit{any function} of $X$ through $Y$. To this end, we define maximal $\alpha$-leakage below.
\begin{definition}[Maximal $\alpha$-Leakage]\label{Def:GeneralLeakge}
	Given a joint distribution $P_{XY}$ on finite alphabets $\mcl X\times\mcl Y$, the maximal $\alpha$-leakage from $X$ to $Y$ is defined as
		\begin{align}
		\label{eq:GealLeak_definition}
		\mcl L_{\alpha}^{\text{max}}(X\to Y)		
		\triangleq\sup_{U- X- Y }\mcl L_{\alpha}(U\to Y)
		\end{align}
	where $\alpha\in[1,\infty]$, $U$ represents any function of $X$ and takes values from an arbitrary finite alphabet. 
\end{definition}

\begin{remark}
	Note that the optimal $P_{\hat{X}}^*$ of the maximization in the denominator of the logarithmic term in \eqref{eq:alphaLeak_definition} minimizes the expectation of the following loss function
	\begin{equation}\label{poly_loss}
	\ell(x,P_{\hat{X}})=\frac{\alpha}{\alpha-1} \big(1-P_{\hat{X}}(x)^{1-\frac{1}{\alpha}}\big),
	\end{equation}
	for each $\alpha\in(1,\infty)$. The limit of the loss function in \eqref{poly_loss} leads to the log-loss (for $\alpha=1$) and 0-1 loss (for $\alpha=\infty$) functions, respectively. In addition, for $\alpha=1$ and $\infty$, the maximal $\alpha$-leakage simplifies to MI and MaxL, respectively. These comments are formalized in the following theorems.
\end{remark}

The following theorem simplifies the expression of the $\alpha$- leakage in \eqref{eq:alphaLeak_definition} by solving the two maximizations in the logarithmic term.
\begin{theorem}\label{Thm:DefEquialentExpression_alphaleakage}
	For $\alpha\in[1,\infty]$, $\alpha$-leakage defined in \eqref{eq:alphaLeak_definition} simplifies to
	\begin{align}\label{eq:alphaLeak_EquivDef}
	\mcl L_{\alpha}(X\to Y)=I_{\alpha}^{\text{A}}(X;Y) \quad \alpha\in[1,\infty].
	\end{align}	
\end{theorem}
The proof hinges on solving the optimal estimations $P^*_{\hat{X}|Y}$ and $P^*_{\hat{X}}$ in \eqref{eq:alphaLeak_definition} for knowing $Y$ or not, respectively, as
\begin{subequations}
	\begin{align}
	P^*_{\hat{X}|Y}(x|y)&=\frac{P_{X|Y}(x|y)^{\alpha}}{\sum_{x }P_{X|Y}(x|y)^{\alpha}}& (x,y)\in \mcl X\times \mcl Y\\
	P^*_{\hat{X}}(x)&=\frac{P_{X}(x)^{\alpha}}{\sum_{x }P_{X}(x)^{\alpha}} &  x\in \mcl X,
	\end{align}
\end{subequations}
and therefore, the logarithm of the ratio in \eqref{eq:alphaLeak_definition} simplifies to A-MI. 
A detailed proof is in Appendix \ref{Proof:DefEquialentExpression_alphaleakage}. 
Making use of the conclusion in Theorem \ref{Thm:DefEquialentExpression_alphaleakage}, the following theorem gives equivalent expressions for the maximal $\alpha$-leakage.
\begin{theorem}\label{Thm:DefEquialentExpression}
	For $\alpha\in[1,\infty]$, the maximal $\alpha$-leakage defined in \eqref{eq:GealLeak_definition} simplifies to\\
	\vspace*{+0.3cm}
	\begin{subequations}
		$\mcl L_{\alpha}^{\text{max}}(X\to Y)$
		\vspace*{-0.3cm}
		\label{eq:GealLeak_EquivDef}
		\begin{empheq}[left={=\empheqlbrace\,}]{align}
		&\sup_{P_{\tilde{X}}}I^{\text{S}}_\alpha(\tilde{X};Y)=\sup_{P_{\tilde{X}}}I_{\alpha}^{\text{A}}(\tilde{X};Y)& \alpha\in(1,\infty]  \label{eq:GealLeak_EquivDef_1infty}\\
		& I(X;Y)  &\alpha=1  \label{eq:GealLeak_EquivDef_1}
		\end{empheq}
	\end{subequations}
where $P_{\tilde{X}}$ has the same support as $P_X$.
\end{theorem}
Note that the maximal $\alpha$-leakage is essentially the Arimoto channel capacity (with a support-set constrained input distribution) for $\alpha\geq 1$ \cite{AlphaMI_Arimoto1975}. This theorem is proved by first applying Theorem \ref{Thm:DefEquialentExpression_alphaleakage} to write the maximal $\alpha$-leakage as
	\begin{align}
	\label{eq:Thm_MaxAlphaLeak_ProofSketch}
	\mcl L_{\alpha}^{\text{max}}(X\to Y)&=\sup_{U-X-Y}I_{\alpha}^{\text{A}}(U;Y)\quad \alpha\in[1,\infty].
	\end{align}
Subsequently, using the facts that A-MI and S-MI have the same supremum \cite[Thm. 5]{alphaMI_verdu} and that S-MI satisfies data processing inequality \cite[Thm. 3]{alphaMI_verdu}, we upper bound the supremum of \eqref{eq:Thm_MaxAlphaLeak_ProofSketch}
by $\sup_{P_{\tilde{X}}}I^{\text{S}}_\alpha(\tilde{X};Y)$, and then, show that the upper bound can be achieved by a specific $U$ with $H(X|U)=0$. A detailed proof can be found in Appendix \ref{Proof:DefEquialentExpression}.

\section{Properties of Maximal $\alpha$-Leakage}\label{Sec:Properties}
In this section, we will prove that maximal $\alpha$-leakage has several properties that one would expect any reasonable leakage measure to have, including: (i) quasi-convexity in the conditional distribution $P_{Y|X}$; (ii) data processing inequalities; and (iii) a composition property.
	
These properties are proved in the following theorem, which makes use of the equivalent form of maximal alpha-leakage found in Theorem \ref{Thm:DefEquialentExpression}, as well as known properties of S-MI from \cite{alphaMI_Sibson1969,ConvexityAlphaMI_Ho,alphaMI_verdu}.
 \begin{theorem}\label{Thm:Geneleak_qusiconvex_nondecreasing_dataprocessing}
 	For $\alpha\in[1,\infty]$, maximal $\alpha$-leakage 
 	\begin{itemize}
 		\item[1.] is quasi-convex in $P_{Y|X}$;
 		\item[2.] is monotonically non-decreasing in $\alpha$;
 		\item[3.] satisfies data processing inequalities: let random variables $X,Y,Z$ form a Markov chain, i.e., $X-Y-Z$, then
 		\begin{subequations}\label{eq:GeneLeak_DataProcessIneq}
 			\begin{align}
 			\mcl L_{\alpha}^{\text{max}}(X\to Z)\leq \mcl L_{\alpha}^{\text{max}}(X\to Y) \label{eq:GeneLeak_DataProcessIneq_XY}\\
 			\mcl L_{\alpha}^{\text{max}}(X\to Z)\leq \mcl L_{\alpha}^{\text{max}}(Y\to Z) \label{eq:GeneLeak_DataProcessIneq_YZ}.
 			\end{align}
 		\end{subequations}
 	\item[4.]
 	satisfies
 	    \begin{align}
 			\mcl L_{\alpha}^{\text{max}}(X\to Y)\geq 0
 		\end{align}with equality if and only if $X$ is independent of $Y$, and 
 		\begin{align}
 		\mcl L_{\alpha}^{\text{max}}(X\to Y)\leq \begin{cases}
 		\log|\mathcal{X}|\quad &\alpha>1\\
 		H(P_X)&\alpha=1 
 		\end{cases}
 		\end{align} with equality if $X$ is a deterministic function of $Y$. 
 	    
     \item[5.] $ \mcl L_{\alpha}^{\text{max}}(X\hspace{-0.05in}\to\hspace{-0.04in}Y)\leq I^{\text{S}}_{\infty}(P_X,P_{Y|X})$ with equality if $P_{Y|X}$ has either 0 or the maximal leakage in Part 4;
     \item[6.] $ \mcl L_{\alpha}^{\text{max}}(X\hspace{-0.05in}\to\hspace{-0.04in}Y)\geq I^{\text{S}}_{\alpha}\left(P^{(\text{u})}_X,P_{Y|X}\right)$, where $P_X^{(\text{u})}$ indicates the uniform distribution of $X$, i.e.,
     \begin{align}
     		\mathsmaller{\mcl L_{\alpha}^{\text{max}}(X\to Y)\geq\frac{\alpha}{\alpha-1}\log\frac{\sum\limits_{y\in\mcl Y}\left(\sum\limits_{x\in \mcl X}P_{Y|X}(y|x)^{\alpha}\right)^{\frac{1}{\alpha}}}{|\mathcal{X}|^{\frac{1}{\alpha}}}.}
     \end{align} The equality holds if either $P_{Y|X}$ is symmetric\footnote{All rows of $P_{Y|X}$ are permutations of other rows, and so are columns.} or $P_{Y|X}$ has 0 leakage.
 	\end{itemize} 
 \end{theorem}
A detailed proof is in Appendix \ref{Proof:Geneleak_qusiconvex_nondecreasing_dataprocessing}.

\begin{remark}
	Note that both MI and MaxL are convex in $P_{Y|X}$ so that $\mcl L^{\text{max}}_{1}(X\to Y)$ and $\mcl L^{\text{max}}_{\infty}(X\to Y)$ are convex in $P_{Y|X}$. 
\end{remark}

Consider two disclosed versions $Y_1$ and $Y_2$ of $X$. The following theorem upper bounds the maximal $\alpha$-leakage to an adversary who has access to both $Y_1$ and $Y_2$ simultaneously.
\begin{theorem}[Composition Theorem]\label{Thm:GeneLeak_CompositionTheory}
	Given a Markov chain $Y_1-X-Y_2$, we have $(\alpha\in [1,\infty])$ 
	\begin{align}
		\mcl L_{\alpha}^{\text{max}}(X\to Y_1,Y_2)\leq \sum_{i\in\{1,2\}}\mcl L_{\alpha}^{\text{max}}(X\to Y_i).
	\end{align}
\end{theorem}
This composition theorem allows composing multiple releases under a total leakage constraint. A detailed proof is in Appendix \ref{proof:Thm:GeneLeak_CompositionTheory}.

\section{Concluding Remarks}\label{Sec:Conclusion Remarks}
Via $\alpha$- and maximal $\alpha$-leakage, we have introduced novel tunable measures for information leakage. These measures can find direct applications in privacy and secrecy problems. The choice of restricting either specific variables or all possible functions of a dataset determines the choice of $\alpha$- and maximal $\alpha$-leakage measures, respectively. Future work includes characterizing privacy-utility tradeoffs for these measures and evaluating existing privacy mappings against these metrics.

\appendices
\section{Proof of Theorem \ref{Thm:DefEquialentExpression_alphaleakage}}\label{Proof:DefEquialentExpression_alphaleakage}
\begin{proof}[\nopunct]
	The expression \eqref{eq:alphaLeak_definition} can be explicitly written as
	\begin{align}	
	&\mcl L_{\alpha}(X\to Y)
	=\lim_{\alpha'\to \alpha}\frac{\alpha'}{\alpha'-1}\nonumber\\
	\label{eq:GealLeak_definition1}
	&\log\left(\frac{\max\limits_{P_{\hat{X}|Y}}\sum\limits_{xy}P_{XY}(xy)\left(P_{\hat{X}|Y}(x|y)\right)^{\frac{\alpha'-1}{\alpha'}}}{\max\limits_{P_{\hat{X}}}\sum\limits_{x}P_X(x)P_{\hat{X}}(x)^{\frac{\alpha'-1}{\alpha'}}}\right).
	\end{align}
	To simplify the expression in \eqref{eq:GealLeak_definition1}, we need to solve the two maximizations in the logarithm. First, we concentrate on the maximization in the denominator of the logarithm in \eqref{eq:GealLeak_definition1} and the one in the numerator can be solved following the same analysis. The maximization in the denominator can be equivalently written as
	\begin{subequations}\label{eq:GealLeak_DefMaxDenominator}
		\begin{align}
		\label{eq:GealLeak_DefMaxDenominator_obj}
		\max_{\substack{P_{\hat{X}}}}\quad &\sum_{x\in\mcl X}P_X(x)P_{\hat{X}}(x)^{1-\frac{1}{\alpha'}}\\
		\label{eq:GealLeak_DefMaxDenominator_const1}
		\text{s.t.}\quad & \sum_{x\in\mcl X}P_{\hat{X}}(x)=1\\
		\label{eq:GealLeak_DefMaxDenominator_const>0}
		& P_{\hat{X}}(x)\geq 0 \quad \text{  for all }x\in \mcl{X} 
		\end{align}
	\end{subequations}
	For $\alpha'\in[1,\infty)$, the problem in \eqref{eq:GealLeak_DefMaxDenominator} is a convex program. Therefore, by using Karush–-Kuhn–-Tucker (KKT) conditions, we obtain the optimal value of \eqref{eq:GealLeak_DefMaxDenominator} as 
	\begin{align}
	\max_{P_{\hat{X}}}\sum_{x\in\mathcal{X}}P_X(x)P_{\hat{X}}(x)^{\frac{\alpha'-1}{\alpha'}}
	=\left(\sum_{x\in\mathcal{X}}P_X(x)^{\alpha'}\right)^{\frac{1}{\alpha'}},
	\end{align}
	with the optimal solution $P^*_{\hat{X}}$ as
	\begin{align}
	\label{eq:GealLeak_DefMaxDenominatorOPTSol}
	P^*_{\hat{X}}(x)=\frac{P_{X}(x)^{\alpha'}}{\sum\limits_{x\in\mcl X}P_{X}(x)^{\alpha'}}\quad \text{for all } x\in\mcl X
	\end{align}	
	Similarly, we attain the optimal solution $P^*_{\hat{X}|Y}$ of the maximization in the numerator of the logarithm in \eqref{eq:GealLeak_definition1} as
	\begin{align}
	\label{eq:GealLeak_DefMaxNumeratorOPTSol}
	P^*_{\hat{X}|Y}(x|y)=\frac{P_{X|Y}(x|y)^{\alpha'}}{\sum\limits_{x\in\mcl X}P_{X|Y}(x|y)^{\alpha'}}
	\end{align}
	for all $x\in\mcl X, y\in\mcl Y$, and therefore, we have
	\begin{align}
	&\max_{P_{\hat{X}|Y}}\sum_{x\in\mathcal{X},y\in\mathcal{Y}}P_{XY}(xy)P_{\hat{X}|Y}(x|y)^{\frac{\alpha'-1}{\alpha'}}\nonumber\\
	=&\sum_{y\in\mcl Y}P_Y(y)\left(\sum_{x\in\mathcal{X}}P_{X|Y}(x|y)^{\alpha'}\right)^{\frac{1}{\alpha'}}.
	\end{align}
	Thus, for $\alpha\in[1,\infty)$, we have 
	\begin{align}
	\label{eq:GealLeak_EquivalentInproof0}
	&\mcl L_{\alpha}(X\to Y)=\nonumber\\
	&\lim_{\alpha'\to \alpha}\frac{\alpha'}{\alpha'-1}\log
	\left(\frac{\sum\limits_{y}P_Y(y)\left(\sum\limits_{x}P_{X|Y}(x|y)^{\alpha'}\right)^{\frac{1}{\alpha'}}}{\left(\sum\limits_{x}P_X(x)^{\alpha'}\right)^{\frac{1}{\alpha'}}}\right),
	\end{align}
	i.e., A-MI of order $\alpha\in[1,\infty)$ in \eqref{eq:Arimoto_MI}.\\
	Note that if $\alpha=\infty$, the optimal solution in \eqref{eq:GealLeak_DefMaxDenominatorOPTSol} is $\frac{0}{0}$. We go back to the expression in \eqref{eq:alphaLeak_definition} and observe that if $\alpha=\infty$, the expression $\mcl L_{\infty}(X\to Y)$ becomes
	\begin{align}\label{eq:GealLeak_EquivalentMax_Inf}
	&\mcl L_{\infty}(X\to Y)\nonumber\\
	=&\log\left(\frac{\max\limits_{P_{\hat{X}|Y}}\sum\limits_{x,y}P_{XY}(xy)P_{\hat{X}|Y}(x|y)}{\max\limits_{P_{\hat{X}}}\sum\limits_{x}P_X(x)P_{\hat{X}}(x)}\right).
	\end{align}
	Since the largest convex combinations is the maximal involved value, the optimal values of the two maximizations in \eqref{eq:GealLeak_EquivalentMax_Inf}
	are 
	\begin{subequations}
		\begin{align}
		&\max_{P_{\hat{X}|Y}}\sum_{xy}P_{XY}(xy)P_{\hat{X}|Y}(x|y)\nonumber\\
		=&\sum_{y} P_Y(y)\max_{x} P_{X|Y}(x|y)\\
		&\max_{P_{\hat{X}}}\sum_{x}P_X(x)P_{\hat{X}}(x)=\max_{x} P_X(x).
		\end{align}
	\end{subequations}
	Therefore, for $\alpha=\infty$, we have
	\begin{align}
	\mcl L_{\infty}(X\to Y)=\log\left(\frac{\sum\limits_{y\in\mcl Y} P_Y(y)\max\limits_{x} P_{X|Y}(x|y)}{\max\limits_{x} P_X(x)}\right),
	\end{align}
	which is exactly the A-MI of order $\infty$. Therefore, $\alpha$-leakage can be equivalently expressed as $I^{\text{A}}_{\alpha}(X;Y)$ for $\alpha\in[1,\infty]$.
\end{proof}

\section{Proof of Theorem \ref{Thm:DefEquialentExpression}}\label{Proof:DefEquialentExpression}
\begin{proof}[\nopunct]
	From Theorem \ref{Thm:DefEquialentExpression_alphaleakage}, we have for $\alpha\in[1,\infty]$,
	\begin{align}
	\label{eq:Inproof_Thm:DefEquialentExpression}
	\mcl L_{\alpha}^{\text{max}}(X\to Y)=\sup_{U- X- Y }I_{\alpha}^{\text{A}}(U;Y).
	\end{align}
	 If $\alpha=1$, we have 
		\begin{align}
			\mcl L_{1}^{\text{max}}(X\to Y)&=\sup_{U- X- Y }I(U;Y)\leq I(X;Y)
		\end{align}
		where the inequality is from data processing inequalities of MI \cite[Thm 2.8.1]{IT_Cover}.\\
	If $\alpha=\infty$,  we have
	 \begin{align}
		\mcl L_{\infty}^{\text{max}}(X\to Y)=\sup_{U- X- Y }\log\frac{\mathsmaller{\sum\limits_{y} P_Y(y)\max\limits_{u} P_{U|Y}(u|y)}}{\mathsmaller{\max\limits_{u} P_U(u)}},
	\end{align}
	which is exactly the expression of MaxL, and therefore, we have \cite[Thm. 1]{MaximalLeakage_Issa2016} 
	\begin{align}
	\mcl L_{\infty}^{\text{max}}(X\to Y)=\log\sum\limits_{y} \max\limits_{x} P_{Y|X}(y|x).
	\end{align}
For $\alpha\in(1,\infty)$, we provide an upper bound for $\mcl L_{\alpha}^{\text{max}}(X\to Y)$, and then, give an achievable scheme as follows.	\\
\textbf{Upper Bound}: 
		We have an upper bound of $\mcl L_{\alpha}^{\text{max}}(X\to Y)$ as
		\begin{subequations}\label{eq:GealLeak_EquivalentInproofConverse}
			\begin{align}
			\label{eq:GealLeak_EquivalentInproofConverse0}
			&\mcl L_{\alpha}^{\text{max}}(X\to Y)\nonumber\\
			=&\sup_{U- X- Y}I_{\alpha}^{\text{A}}(U;Y)\\
			\label{eq:GealLeak_EquivalentInproofConverse2}
			\leq &\sup_{P_{\tilde{X}|\tilde{U}}:P_{\tilde{X}|\tilde{U}}\left(\cdot|u\right)\ll P_{X}} \sup_{P_{\tilde{U}}} I_{\alpha}^{\text{A}}(\tilde{U};Y)\\
			\label{eq:GealLeak_EquivalentInproofConverse3}
			= &\sup_{P_{\tilde{X}|\tilde{U}}:P_{\tilde{X}|\tilde{U}}\left(\cdot|u\right)\ll P_{X}} \sup_{P_{\tilde{U}}} I_{\alpha}^{\text{S}}(\tilde{U};Y)\\
			\label{eq:GealLeak_EquivalentInproofConverse4}
			= &\sup_{P_{\tilde{X}}\ll P_{X}} I_{\alpha}^{\text{S}}(\tilde{X};Y)\\
			\label{eq:GealLeak_EquivalentInproofConverse5}
			= &\sup_{P_{\tilde{X}}\ll P_{X}} I_{\alpha}^{\text{A}}(\tilde{X};Y)
			\end{align}
		\end{subequations}
		where $P_{\tilde{X}}\ll P_{X}$ means the alphabet of $P_{\tilde{X}}$ is a subset of that of $P_{X}$. The inequality in \eqref{eq:GealLeak_EquivalentInproofConverse2} holds because the supremum of A-MI over all $P_{\tilde{U},\tilde{X}}$ on $\mcl U\times \mcl X$ is no less than that (in \eqref{eq:GealLeak_EquivalentInproofConverse0}) over these $P_{U,X}$ constrained by the $P_X$. The equations in \eqref{eq:GealLeak_EquivalentInproofConverse3} and \eqref{eq:GealLeak_EquivalentInproofConverse5} result from that A-MI and S-MI of order $\alpha>0$ have the same supremum \cite[Thm. 5]{alphaMI_verdu}; and \eqref{eq:GealLeak_EquivalentInproofConverse4} obeys the data processing inequalities \cite[Thm. 3]{alphaMI_verdu}.\\
\textbf{Lower bound}: 
		We lower bound \eqref{eq:Inproof_Thm:DefEquialentExpression} by consider a random variable $U$ such that $U-X-Y$ is a Markov chain and $H(X|U)=0$. Specifically, let the alphabet $\mcl U$ consist of $\mcl U_x$, a collection of $U$ mapped to a $x\in \mcl X$, i.e.,
		$\mcl U=\cup_{x\in\mcl X} \mcl U_x $ with $U=u\in \mcl U_x$ if and only if $X=x$.
		Therefore, for the specific variable $U$, we have
		\begin{align}
		\label{eq:GealLeak_EquivalentInproofAchieval0}
		P_{Y|U}(y|u)&=\begin{cases}
		P_{Y|X}(y|x) \quad &\text{ for all } u\in  \mcl U_x\\
		0                  &\text{ otherwise}.
		\end{cases}
		\end{align}
		Construct a probability distribution $P_{\tilde{X}}$ over $\mathcal{X}$ from $P_U$ as
		\begin{align}
		\label{eq:GealLeak_EquivalentInproofAchievalConstructPX}
		P_{\tilde{X}}(x)=\frac{\sum_{u\in\mathcal{U}_x}P_U^{\alpha}(u)}{\sum_{x\in\mcl X}\sum_{u\in\mathcal{U}_x}P_U^{\alpha}(u)}  \quad \text{ for all }  x\in \mcl X.
		\end{align} Thus, 
		\begin{align*}
		&I_{\alpha}^{\text{A}}(U;Y)\nonumber\\
		=&\frac{\alpha}{\alpha-1}\log\frac{\sum\limits_{y\in\mcl Y}\left(\sum\limits_{x\in\mcl X}\sum\limits_{u\in\mathcal{U}_x}P_{Y|U}(y|u)^{\alpha}P_{U}(u)^{\alpha}\right)^{\frac{1}{\alpha}}}{\left(\sum\limits_{x\in\mcl X}\sum\limits_{u\in\mathcal{U}_x}P_U(u)^{\alpha}\right)^{\frac{1}{\alpha}}}
		\end{align*}
		   \begin{align*}
		   =&\frac{\alpha}{\alpha-1}\log\frac{\sum\limits_{y\in\mcl Y}\left(\sum\limits_{x\in\mcl X}P_{Y|X}(y|x)^{\alpha}\sum\limits_{u\in\mathcal{U}_x}P_{U}(u)^{\alpha}\right)^{\frac{1}{\alpha}}}{\left(\sum\limits_{x\in\mcl X}\sum\limits_{u\in\mathcal{U}_x}P_U(u)^{\alpha}    \right)^{\frac{1}{\alpha}}}	\\
			=&\frac{\alpha}{\alpha-1}\log\left(\sum_{y\in\mcl Y}\left(\sum_{x\in\mcl X}P_{Y|X}(y|x)^{\alpha}P_{\tilde{X}}(x)^{\alpha}\right)^{\frac{1}{\alpha}}\right)\\
			=&I_{\alpha}^{\text{S}}(\tilde{X};Y)
			\end{align*}
		Therefore, 
		\begin{subequations}\label{eq:GealLeak_EquivalentInproofAchievable}
			\begin{align}
			\mcl L_{\alpha}^{\text{max}}(X\to Y) =&\sup_{U-X-Y} I_{\alpha}^{\text{A}}(U;Y)\nonumber\\
			\geq &\sup_{U:U-X-Y,H(X|U)=0} I_{\alpha}^{\text{A}}(U;Y)\\
			\label{eq:GealLeak_EquivalentInproofAchieval3}
			=&\sup_{P_{\tilde{X}}\ll P_X}I_{\alpha}^{\text{S}}(\tilde{X};Y),
			\end{align}
		\end{subequations}	
	where \eqref{eq:GealLeak_EquivalentInproofAchieval3} is because for any $P_{\tilde{X}}\ll P_X$, it can be obtained through \eqref{eq:GealLeak_EquivalentInproofAchievalConstructPX} by appropriately choosing $P_U$.
	Therefore, combining \eqref{eq:GealLeak_EquivalentInproofConverse} and \eqref{eq:GealLeak_EquivalentInproofAchievable}, we obtain \eqref{eq:GealLeak_EquivDef_1infty}. 
\end{proof}

\section{Proof of Theorem \ref{Thm:Geneleak_qusiconvex_nondecreasing_dataprocessing}}\label{Proof:Geneleak_qusiconvex_nondecreasing_dataprocessing}
\begin{proof}[\nopunct]
    \textbf{The proof of part 1}: We know that for $\alpha\geq 1$, $I^{\text{S}}_{\alpha}(X;Y)$ is quasi-convex $P_{Y|X}$ for given $P_X$ \cite[Thm. 2.7.4]{IT_Cover}, \cite[Thm. 10]{ConvexityAlphaMI_Ho}. In addition, the supreme of a set of quasi-convex functions is also quasi-convex, i.e., let function $f(a,b)$ is quasi-convex in $b$, such that $\sup_a f(a,b)$ is also quasi-convex in $b$ \cite{boydconvex}. Therefore, the maximal $\alpha$-leakage in \eqref{eq:GealLeak_EquivDef} is quasi-convex $P_{Y|X}$ for given $P_X$.\\
    \textbf{The proof of part 2}: 
	Let $\beta>\alpha\geq1$, and $P_{X\alpha}^*=\arg \sup_{P_X} I^{\text{S}}_{\alpha}(P_X,P_{Y|X})$ for given $P_{Y|X}$, such that
	\begin{subequations}
		\begin{align}
			\mcl L_{\alpha}^{\text{max}}(X\to Y)&= I^{\text{S}}_{\alpha}(P_{X\alpha}^*,P_{Y|X})\\
			\label{eq:GeneLeak_Property1inProof4}
			& \leq I^{\text{S}}_{\beta}(P_{X\alpha}^*,P_{Y|X})\\
			\label{eq:GeneLeak_Property1inProof5}
			& \leq \sup_{P_X} I^{\text{S}}_{\beta}(P_X,P_{Y|X})\\
			&=\mcl L^{\text{max}}_{\beta}(X\to Y)
		\end{align}
	\end{subequations}
	where \eqref{eq:GeneLeak_Property1inProof4} results from that $I^{\text{S}}_{\alpha}$ is non-decreasing in $\alpha$ for $\alpha>0$ \cite[Thm. 4]{ConvexityAlphaMI_Ho}, and the equality in \eqref{eq:GeneLeak_Property1inProof5} holds if and only if $P_{X\alpha}^*=\arg \sup_{P_X} I_{\beta}(P_X,P_{Y|X})$.\\	
	\textbf{The proof of part 3}: 
     Let random variables $X$, $Y$ and $Z$ form the Markov chain $X-Y-Z$. Making use of that S-MI of order $\alpha>1$ satisfies data processing inequalities \cite[Thm. 3]{alphaMI_verdu}, i.e., 
	\begin{subequations}
		\begin{align}
		I^{\text{S}}_{\alpha}(X; Z)\leq I^{\text{S}}_{\alpha}(X; Y) \label{eq:DPInq_inproof01}\\
		I^{\text{S}}_{\alpha}(X; Z)\leq I^{\text{S}}_{\alpha}(Y; Z) \label{eq:DPInq_inproof02},
		\end{align}
	\end{subequations} 
    we prove that maximal $\alpha$-leakage satisfies data processing inequalities as follows.\\
	We first prove \eqref{eq:GeneLeak_DataProcessIneq_XY}. Let $P^*_X=\arg\sup_{P_X} I^{\text{S}}_{\alpha}(P_X,P_{Z|X})$. For the Markov chain $X-Y-Z$, we have
	\begin{subequations}
		\begin{align}
		\mcl L_{\alpha}^{\text{max}}(X\to Z)&=I^{\text{S}}_{\alpha}(P^*_X,P_{Z|X}) \label{eq:DPInq_inproof1}\\
		&\leq I^{\text{S}}_{\alpha}(P^*_X,P_{Y|X}) \label{eq:DPInq_inproof2}\\
		&\leq \sup_{P_X} I^{\text{S}}_{\alpha}(P_X,P_{Y|X}) \label{eq:DPInq_inproof3}\\
		&=\mcl L_{\alpha}^{\text{max}}(X\to Y) \label{eq:DPInq_inproof4}
		\end{align}
	\end{subequations}
	where the inequality in \eqref{eq:DPInq_inproof2} results from \eqref{eq:DPInq_inproof01}. Similarly, the inequality in \eqref{eq:GeneLeak_DataProcessIneq_YZ} can be proved directly from \eqref{eq:DPInq_inproof02}.\\	
    \textbf{The proof of part 4}: 
	For $\alpha\in(1,\infty]$, referring to \eqref{eq:Sibson_MI} and \eqref{eq:GealLeak_EquivDef_1infty} we have 
	\begin{subequations}
		\begin{align}
		&\mcl L_{\alpha}^{\text{max}}(X\to Y)\nonumber\\
		=&\sup_{P_X}	\frac{\alpha}{\alpha-1}\log \sum_{y}\left(\sum_{x}P_X(x)P_{Y|X}(y|x)^{\alpha}\right)^{\frac{1}{\alpha}}\\
		\label{eq:alphaLeak_SpecialMechanism_Inproof1}
		\geq & \sup_{P_X}	\frac{\alpha}{\alpha-1}\log \sum_{y}\bigg(\sum_{x}P_X(x)P_{Y|X}(y|x)\bigg)^{\frac{\alpha}{\alpha}}\\
		=& \sup_{P_X}	\frac{\alpha}{\alpha-1}\log 1=0,
		\end{align}
	\end{subequations}	
	where \eqref{eq:alphaLeak_SpecialMechanism_Inproof1} results from applying Jensen’s inequality to the convex function $f: t\to t^{\alpha}$ ($t\geq 0$), such that the equality holds if and only if given any $y\in\mcl Y$, $P_{Y|X}(y|x)$ are the same for all $x\in \mcl X$, such that
	\begin{align}
	P_{Y|X}(y|x)=P_Y(y)\quad x\in \mcl X, y\in\mcl Y
	\end{align} which means $X$ and $Y$ are independent, i.e., $P_{Y|X}$ is a rank-1 row stochastic matrix. 
	For $\alpha=1$, we have
	\begin{align}
	\mcl L^{\text{max}}_{1}(X\to Y)=I(X;Y)\geq 0,
	\end{align}
	with equalities if and only if $X$ is independent of $Y$ \cite{IT_Cover}. \\	
	Let $P_{X\Leftarrow Y}$ be an conditional probability matrix with only one non-zero entry in each column, and indicate the only non-zero entries by $x_y$, i.e., $x_y=\arg_x P_{X\Leftarrow Y}(y|x)>0$ for all $y\in\mcl Y$. 
	For $\alpha=\infty$, we have 
	\begin{align}
	\mcl L^{\text{max}}_{\infty}(P_{X\Leftarrow Y})
	&=\log \sum_{y\in\mcl Y}P_{X\Leftarrow Y}(y|x_y)=\log |\mcl X|,
	\end{align}
	which is exactly the upper bound of MaxL \cite[Lem. 1]{MaximalLeakageHT_Liao2017} and absolutely an upper bound of maximal $\alpha$ leakage due to its monotonicity in $\alpha$.\\
	For $\alpha\in(1,\infty)$, from \eqref{eq:Sibson_MI} and \eqref{eq:GealLeak_EquivDef_1infty} we have
	\begin{subequations}
		\begin{align}
		&\mcl L_{\alpha}^{\text{max}}(P_{X\Leftarrow Y})\nonumber\\
		=&\sup_{P_X} \frac{\alpha}{\alpha-1}\log \sum_{y\in\mathcal{Y}}\left(P_X^{\frac{1}{\alpha}}(x_y)P_{X\Leftarrow Y}(y|x_y)\right)\\
		\label{eq:GeneLeak_LemmaSpecialMech_proof1}
		= &\sup_{P_X} \frac{\alpha}{\alpha-1}\log \sum_{x\in\mathcal{X}}P_X^{\frac{1}{\alpha}}(x);
		\end{align}
	\end{subequations}	
	in addition, since the function maximized in \eqref{eq:GeneLeak_LemmaSpecialMech_proof1} is symmetric and concave in $P_X$, it is Schur-concave in $P_X$, and therefore, the optimal distribution of $X$ achieving the supreme in \eqref{eq:GeneLeak_LemmaSpecialMech_proof1} is uniform. Thus, 
	\begin{align}
	\mcl L_{\alpha}^{\text{max}}(P_{X\Leftarrow Y})=\log |\mathcal{X}|\quad \text{ for } \alpha\in(1,\infty) .
	\end{align}
	For $\alpha=1$, referring to \eqref{eq:GealLeak_EquivDef_1} we have  
	\begin{subequations}
		\begin{align}
		&\mcl L^{\text{max}}_{1}(X\to Y)\nonumber\\
		=&\sum_{y\in\mcl Y}P_X(x_y)P_{X\Leftarrow Y}(y|x_y)\log \frac{P_{X\Leftarrow Y}(y|x_y)}{P_X(x_y)P_{X\Leftarrow Y}(y|x_y)}\\
		=&\sum_{y\in\mcl Y}P_X(x_y)P_{X\Leftarrow Y}(y|x_y)\log \frac{1}{P_X(x_y)}\\
		=&\sum_{x\in\mcl X}P_X(x)\log \frac{1}{P_X(x)}=H(P_X),
		\end{align}
	\end{subequations}
	which is exactly the upper bound of $I(X;Y)$.\\
	Therefore, if $X$ is a deterministic function of $Y$, maximal $\alpha$-leakage achieves its maximal value $\log|\mathcal{X}|$ for $\alpha>1$, and $H(P_X)$ for $\alpha=1$.\\
    \textbf{The proof of part 5}: 
	The upper bound is directly from the fact that maximal $\alpha$-leakage is non-decreasing in $\alpha$. In addition, from the results in part 4, we know that if $P_{Y|X}$ has 0 or the maximal leakage in in part 4, the upper bound is tight.\\  
	\textbf{The proof of part 6}:
	Given $P_{Y|X}$, the lower bound is actually the S-MI of order $\alpha$ for the uniform distribution of $X$. Due to the concavity of $I^{\text{S}}_{\alpha}(P_X,P_{Y|X})$ ($\alpha\geq 1$) in $P_X$ \cite[Thm. 8]{ConvexityAlphaMI_Ho} \footnote{The concavity of $I^{\text{S}}_{\alpha}(P_X,P_{Y|X})$ is based on the fact that a conditional R{\'e}nyi divergence is concave in $P_X$ \cite{ConvexityAlphaMI_Ho}.}, we know that $I^{\text{S}}_{\alpha}(P_X,P_{Y|X})$ is Schur concave in $P_X$ for any symmetric $P_{Y|X}$. Therefore the uniform distribution of $X$ maximizes \eqref{eq:GealLeak_EquivDef_1infty} 
	and its S-MI is exactly the maximal $\alpha$-leakage \cite[Col. 9]{ConvexityAlphaMI_Ho}\footnote{Let $f(\mathbf{x})$ be a function which is Schur concave in a vector variable $\mathbf{x}\in \mathbb{R}^n$, $\mathbf{x}_1$ and $\mathbf{x}_2$ be two decreasing-ordered vectors in the domain of $f(\mathbf{x})$. If $\mathbf{x}_1$ majors $\mathbf{x}_2$, i.e., $\sum_{1}^{k}x_{1i}\geq \sum_{1}^{k}x_{2i}$ (for all $k\leq n$) and $\sum_{1}^{n}x_{1i}= \sum_{1}^{n}x_{2i}$, then $f(\mathbf{x}_1)\leq f(\mathbf{x}_2)$.}.
	The $P_{Y|X}$ in part 4 with zero leakage make the lower bound tight.
\end{proof}

\section{Proof of Theorem \ref{Thm:GeneLeak_CompositionTheory}}\label{proof:Thm:GeneLeak_CompositionTheory}
\begin{proof}[\nopunct]
	Let $\mcl Y_1$ and $\mcl Y_2$ be the alphabets of $Y_1$ and $Y_2$, respectively. For any $(y_1,y_2)\in \mcl Y_1\times \mcl Y_2$, due to the Markov chain $Y_1-X-Y_2$, the corresponding entry of the conditional probability matrix of $(Y_1,Y_2)$ given $X$ is
	\begin{align}
		P(y_1,y_2|x)=P(y_1|x)P(y_2|x,y_1)=P(y_1|x)P(y_2|x).
	\end{align}
	Therefore, for $\alpha\in(1,\infty)$
	\begin{subequations}
		\begin{align}
			&\mcl L_{\alpha}^{\text{max}}(X\to Y_1,Y_2)\nonumber\\
			=&\sup_{P_X} \frac{\alpha}{\alpha-1}\log \sum_{y_1,y_2\in \mcl Y_1\times \mcl Y_2}\nonumber\\
			&\quad\left(\sum_{x\in\mathcal{X}}P_X(x)P_{Y_1,Y_2|X}(y_1,y_2|x)^{\alpha}\right)^{\frac{1}{\alpha}}\\
			=&\sup_{P_X} \frac{\alpha}{\alpha-1}\log \sum_{y_1,y_2\in \mcl Y_1\times \mcl Y_2}\nonumber\\
			\label{eq:alphaLeakage_CompostionTheoremProof_0}
			&\quad\left(\sum_{x\in\mathcal{X}}P_X(x)P_{Y_1|X}(y_1|x)^{\alpha}P_{Y_2|X}(y_2|x)^{\alpha}\right)^{\frac{1}{\alpha}}.
		\end{align}
	\end{subequations}		
	Let $K(y_1)=\sum_{x\in\mathcal{X}}P_X(x)P_{Y_1|X}(y_1|x)^{\alpha}$, for all $y_1\in\mcl Y_1$, such that we can construct a set of distributions over $\mcl X$ as 
	\begin{align}
		P_{\tilde{X}}(x|y_1)=\frac{P_X(x)P_{Y_1|X}(y_1|x)^{\alpha}}{K(y_1)}.
	\end{align} Therefore, from \eqref{eq:alphaLeakage_CompostionTheoremProof_0}, $\mcl L_{\alpha}^{\text{max}}(X\to Y_1,Y_2)$ can be rewritten as
		\begin{align}
		&\mcl L_{\alpha}^{\text{max}}(X\to Y_1,Y_2)\nonumber\\
		=&\sup_{P_X} \frac{\alpha}{\alpha-1}\log \sum_{y_1,y_2\in \mcl Y_1\times \mcl Y_2}\left(\sum_{x\in\mathcal{X}}K(y_1)P_{\tilde{X}}(x|y_1)\right.\nonumber\\
		&  P_{Y_2|X}(y_2|x)^{\alpha}\bigg)^{\frac{1}{\alpha}}\displaybreak[0]\\
		=&\sup_{P_X} \frac{\alpha}{\alpha-1}\log \sum_{y_1,y_2\in \mcl Y_1\times \mcl Y_2}\left(\sum_{x\in\mathcal{X}}P_X(x)\right.\nonumber\\
		&P_{Y_1|X}(y_1|x)^{\alpha}\bigg)^{\frac{1}{\alpha}}\left(\sum_{x\in\mathcal{X}} P_{\tilde{X}}(x|y_1)P_{Y_2|X}(y_2|x)^{\alpha}\right)^{\frac{1}{\alpha}}\\
		= &\sup_{P_X} \frac{\alpha}{\alpha-1}\log \mathlarger{\sum}_{y_1\in \mcl Y_1}\small{\left(\sum_{x\in\mathcal{X}}P_X(x)P_{Y_1|X}(y_1|x)^{\alpha}\right)^{\frac{1}{\alpha}}}\nonumber\\
		&\small{\sum_{y_2\in \mcl Y_2}\left(\sum_{x\in\mathcal{X}} P_{\tilde{X}}(x|y_1)P_{Y_2|X}(y_2|x)^{\alpha}\right)^{\frac{1}{\alpha}}}
		\end{align}
		\begin{align}
		\leq & \mathsmaller{\sup\limits_{P_X} \frac{\alpha}{\alpha-1}\log \left(\sum\limits_{y_1\in \mcl Y_1}\left(\sum\limits_{x\in\mathcal{X}}P_X(x)P_{Y_1|X}(y_1|x)^{\alpha}\right)^{\frac{1}{\alpha}}\right.}\nonumber\\
		\label{eq:alphaLeakage_CompostionTheoremProof_1}
		&\mathsmaller{\left.\max\limits_{y_1\in\mcl Y_1}\sum\limits_{y_2\in \mcl Y_2}\left(\sum\limits_{x\in\mathcal{X}} P_{\tilde{X}}(x|y_1)P_{Y_2|X}(y_2|x)^{\alpha}\right)^{\frac{1}{\alpha}}\right)}\\			
		=&\mathsmaller{\sup\limits_{P_X} \frac{\alpha}{\alpha-1}\log \left(\sum\limits_{y_1\in \mcl Y_1}\left(\sum\limits_{x\in\mathcal{X}}P_X(x)P_{Y_1|X}(y_1|x)^{\alpha}\right)^{\frac{1}{\alpha}}\right.}\nonumber\\
		\label{eq:alphaLeakage_CompostionTheoremProof_2}
		&\mathsmaller{\left.\sum\limits_{y_2\in \mcl Y_2}\left(\sum\limits_{x\in\mathcal{X}} P_{\tilde{X}}(x|y_1^*)P_{Y_2|X}(y_2|x)^{\alpha}\right)^{\frac{1}{\alpha}}\right)}\\				
		\leq &\mathsmaller{\sup\limits_{P_X} \frac{\alpha}{\alpha-1}\log \left(\sum\limits_{y_1\in \mcl Y_1}\left(\sum\limits_{x\in\mathcal{X}}P_X(x)P_{Y_1|X}(y_1|x)^{\alpha}\right)^{\frac{1}{\alpha}}\right.}\nonumber\\
		\label{eq:alphaLeakage_CompostionTheoremProof_3}
		&\mathsmaller{+\sup\limits_{P_{\tilde{X}}}\frac{\alpha}{\alpha-1}\log\sum\limits_{y_2\in \mcl Y_2}\left(\sum\limits_{x\in\mathcal{X}} P_{\tilde{X}}(x)P_{Y_2|X}(y_2|x)^{\alpha}\right)^{\frac{1}{\alpha}}}\\
		=&\mcl L_{\alpha}^{\text{max}}(X\to Y_1)+\mcl L_{\alpha}^{\text{max}}(X\to Y_2).
		\end{align}
	where $y_1^*$ in \eqref{eq:alphaLeakage_CompostionTheoremProof_2} is the optimal $y_1$ achieving the maximum in \eqref{eq:alphaLeakage_CompostionTheoremProof_1}. Therefore, the equality in \eqref{eq:alphaLeakage_CompostionTheoremProof_1} holds if and only if, for all $y_1\in \mcl Y_1$,
	\begin{align}
		&\sum_{y_2\in \mcl Y_2}\left(\sum_{x\in\mathcal{X}} P_{\tilde{X}}(x|y_1)P_{Y_2|X}(y_2|x)^{\alpha}\right)^{\frac{1}{\alpha}}\nonumber\\
		=&\sum_{y_2\in \mcl Y_2}\left(\sum_{x\in\mathcal{X}} P_{\tilde{X}}(x|y_1^*)P_{Y_2|X}(y_2|x)^{\alpha}\right)^{\frac{1}{\alpha}};
	\end{align}
	and the equality in \eqref{eq:alphaLeakage_CompostionTheoremProof_3} holds if and only if the optimal solutions $P_X^*$ and $P_{\tilde{X}}^*$ of the two maximizations in \eqref{eq:alphaLeakage_CompostionTheoremProof_3} satisfy, for all $x\in\mcl X$,
	\begin{align}
		P_{\tilde{X}}^*(x)=\frac{P_X^*(x)P_{Y_1|X}^{\alpha}(y_1^*|x)}{\sum_{x\in\mathcal{X}}P_X(x)P_{Y_1|X}^{\alpha}(y_1^*|x)}.
	\end{align}
	Now we consider $\alpha=1$. For $Y_1-X-Y_2$, we have
	\begin{align}
		I(Y_2;X|Y_1)\leq I(Y_2;X).
	\end{align}
	From Theorem \ref{Thm:DefEquialentExpression}, there is
	\begin{subequations}
		\begin{align}
		&\mcl L^{\text{max}}_{1}(X\to Y_1,Y_2)\nonumber\\
		=&I(X;Y_1)+I(X;Y_2|Y_1)\\
		\leq &I(X;Y_1)+I(X;Y_2)\\
		=& \mcl L^{\text{max}}_{1}(X\to Y_1)+\mcl L^{\text{max}}_{1}(X\to Y_2).
		\end{align}
	\end{subequations}
	For $\alpha=\infty$, we also have
	\begin{subequations}
		\begin{align}
		&\mcl L^{\text{max}}_{\infty}(X\to Y_1,Y_2)\nonumber\\
		=&\log \sum_{y_1,y_2\in \mcl Y_1\times \mcl Y_2}\max_{x\in\mcl X} P(y_1|x)P(y_2|x)\\
		\leq &\log \sum_{y_1,y_2\in \mcl Y_1\times \mcl Y_2}\left(\max_{x\in\mcl X} P(y_1|x)\right)\left(\max_{x\in\mcl X} P(y_2|x)\right)\\
		=&\log \sum_{y_1\in \mcl Y_1}\max_{x\in\mcl X} P(y_1|x)+\log \sum_{y_2\in \mcl Y_2}\max_{x\in\mcl X} P(y_2|x)\\
		=&	\mcl L^{\text{max}}_{\infty}(X\to Y_1)+	\mcl L^{\text{max}}_{\infty}(X\to Y_2).
		\end{align}
	\end{subequations}
	
\end{proof}

\section*{Acknowledgment}
The authors would like to thank Prof. Vincent Y. F. Tan from National University of Singapore 
for many valuable discussions.

\bibliographystyle{IEEEtran}
\bibliography{JL_References}
\end{document}